\newcommand{\mytitle}[1]{ \begin{center}\bfseries \Huge {#1} 
    \normalsize \mdseries \end{center} \vspace{1cm}}
\newcommand{\myauthor}[2]{\begin{center}{#1} \vspace{1ex} \\ {#2}\end{center}
    \vspace{2cm}}

\newcommand{\myabstract}{\begin{center}\bfseries Abstract \mdseries\end{center}}

\newcommand{\refeq}[1]{(\ref{#1})}

\newcommand{\pk}[1]{{\rm #1}}

\newcommand{\bmath}[1]{\mbox{\boldmath${#1}$\unboldmath}}

\newcommand{\ip}[2]{\ensuremath{{#1}\!\cdot\!{#2}}}
\newcommand{\cross}[2]{\ensuremath{{#1}\bmath{\times}{#2}}}

\newcommand{\dkub}[1]{d^{3}\!{#1}}

\newcommand{\expup}[1]{e^{#1}}

\newcommand{\nuclide}[3]{\ensuremath{\rule{0em}{1.5ex}^{#1}_{#2}{\pk{#3}}}}

\documentclass[twoside,12pt,a4paper]{article}
\usepackage{graphics}

\begin{document}

\setlength{\unitlength}{1mm}

\mytitle{Threshold Production of \\ \vspace{1ex} the Hypertriton}

\myauthor{Anders G\aa rdestig\footnote{e-mail: grdstg@tsl.uu.se}} 
    {Division of Nuclear Physics, Uppsala University,
    \\ Box 535, S-751 21 Uppsala, Sweden}

\myabstract

The cross section for threshold production of the hypertriton in the reaction 
$\pk{pd}\rightarrow\nuclide{3}{\Lambda}{\! H}\pk{K}^{+}$ is calculated in a 
two-step model
and compared to the break-up process 
$\pk{pd}\rightarrow\pk{d}\Lambda\pk{K}^{+}$. The
latter process is shown to be dominant already at 2 MeV above threshold. The
amplitude squared at threshold for the 
$\pk{pd}\rightarrow\nuclide{3}{\Lambda}{\! H}\pk{K}^{+}$ reaction is
 $|f|^{2}=1.0\ \rm nb/sr$.

\clearpage

\section{Introduction}
\label{sec:intro}

One way to gain information about the hyperon-nucleon interaction is to
study bound states of the hyperon, \emph{i.e.}, hypernuclei. The most
suitable candidate for such a study would be the lightest of the hypernuclei,
$\nuclide{3}{\Lambda}{\! H}$, since the theoretical treatment becomes
considerably easier for few-body systems (no collective effects that screens
the perturbation caused by the hyperon). However, the hypertriton is very weakly
bound ($\pk{E_{B}}$= 0.13 MeV), which makes the distinction between the
bound and the scattering state difficult. Because of the weak binding, the decay
properties of the hypertriton are essentially those of the hyperon itself.

The purpose of the present study is to calculate the cross section for the
 process
\begin{equation}
    \pk{pd}\rightarrow\nuclide{3}{\Lambda}{\! H}\pk{K}^{+}
\end{equation}
near threshold ($T_{\pk{p}}^{\pk{lab}}=1126.517\ {\rm MeV}$).
In the similar reaction
$\pk{pd}\rightarrow\nuclide{3}{}{He}\,\eta$, the
two-step model of \cite{laget} has been successful in
explaining the experimental data. Hence, the same mechanism is employed here 
for the hypertriton production. The fundamental
advantage of the two-step model is that it allows the exchanged momentum to be
shared between the two nucleons of the deuteron, thereby increasing the reaction
probability. In addition, the kinematics is
miraculous in the sense that the subprocesses can be considered as real, which
simplifies the calculations.

The concept of a transition matrix, with a definition close to that of
Kondratyuk \emph{et al.}~\cite{konuz} and Komarov \emph{et al.}~\cite{kul},
gives a somewhat different formulation than the one of~\cite{FW}. 
In \cite{kul}, the
cross section for hypertriton production is calculated for several angles
and preferently at energies far from threshold, while this study is devoted to
the behavior near to threshold and the relation to the corresponding
 break-up reaction. 

The index $\tau$ will in this paper stand for the 
($\Lambda$-)hypertriton.


\section{Bound final state}
\label{sec:bound}

\begin{figure}[ht]

\begin{picture}(120,70)(-30,-20)


    \put(-30,-2){\line(1,0){23.81}}
        \put(-15,5){\makebox(0,0){$\pk{d}$}}
        \put(-15,-5){\makebox(0,0){$p_{\pk{d}}$}}
    \put(-30,2){\line(1,0){23.81}}
    \put(0,0){\circle{13}} \put(0,0){\makebox(0,0){B}}
    \put(6.5,0){\line(1,0){47}}
        \put(30,3){\makebox(0,0){$\pk{n}$}}
        \put(30,-3){\makebox(0,0){$\frac{1}{2}p_{\pk{d}}+q$}}
    \put(60,0){\circle{13}} \put(60,0){\makebox(0,0){C}}
    \multiput(66.5,0)(6.22,0){7}{\line(1,0){4.3}}
        \put(85,3){\makebox(0,0){$\pk{K}^{+}$}}
        \put(85,-3){\makebox(0,0){$p_{\pk{K}}$}}


    \put(-30,40){\line(1,0){43.5}}
        \put(-5,43){\makebox{$\pk{p}$}} 
        \put(-5,37){\makebox(0,0){$-p_{\pk{d}}$}}
    \put(20,40){\circle{13}} \put(20,40){\makebox(0,0){A}}
    \put(26.195,42){\line(1,0){47.61}}
        \put(50,45){\makebox(0,0){$\pk{d'}$}}
        \put(50,35){\makebox(0,0){$-(1\!-\!\gamma)p_{\pk{K}}-k$}}
    \put(26.195,38){\line(1,0){47.61}}
    \put(80,40){\circle{13}} \put(80,40){\makebox(0,0){D}}
    \put(85.82,43){\line(1,0){24.18}}
        \put(95,46){\makebox(0,0){$\nuclide{3}{\Lambda}{\! H}$}}
    \put(86.5,40){\line(1,0){23.5}}
        \put(95,34){\makebox(0,0){$-p_{\pk{K}}$}}
    \put(85.82,37){\line(1,0){24.18}}


    \put(2.907,5.814){\line(1,2){14.186}}
        \put(10,20){\makebox(0,0)[br]{$\pk{p'}$}} 
        \put(11,20){\makebox(0,0)[tl]{$\frac{1}{2}p_{\pk{d}}-q$}}
    \multiput(24.596,35,404)(4.401,-4.401){7}{\line(1,-1){3.7}}
        \put(40,20){\makebox(0,0)[tr]{$\pi^{+}$}}
        \put(41,20){\makebox(0,0)[bl]{$p_{\pi}$}}
    \put(62.907,5.814){\line(1,2){14.186}}
        \put(69,20){\makebox(0,0)[br]{$\Lambda$}}
        \put(70,20){\makebox(0,0)[tl]{$-\gamma p_{\pk{K}}+k$}}

\end{picture}

    \caption{Feynman diagram of the two-step model for the 
    $\pk{pd} \rightarrow \protect\nuclide{3}{\Lambda}{\! H}\pk{K}^{+}$ reaction,
    defining the various momenta. The partition of the hypertriton momenta
    is weighted by
    $\gamma = \pk{m}_{\!\Lambda}/(\pk{m}_{\pk{d}}+\pk{m}_{\!\Lambda})=0.373$.
    The diagram for the $\pi^{0}$ case follows upon exchanging $\pk{p'}$ and
    $\pk{n}$.}
    \label{fig:feynbound}
\end{figure}
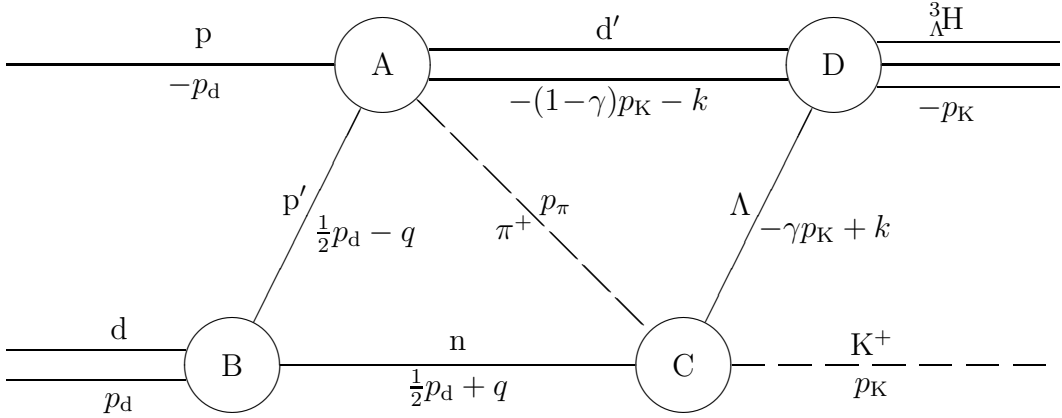

For the case of a bound state between the deuteron and the lambda hyperon, a
two step model like that of \cite{laget} is used. The Feynman diagram for this
 reaction is shown in Fig.~\ref{fig:feynbound}. The trick used in calculating
 the cross section for this diagram is to consider the processes of the $A$ and
$C$ vertices to be almost real. This is accomplished by putting the intermediate
deuteron and neutron on their mass shells.

\subsection{Cross section for bound final state}

With the normalization of Bjorken and Drell \cite{BD} the 
cross section in the overall c.m.\ system can be written
\begin{equation}
    \left[ \frac{d\sigma}{d\Omega}
    (\pk{pd} \rightarrow\nuclide{3}{\Lambda}{\! H}\pk{K}^{+}) \right]_{\pk{CM}}
    = \frac{\pk{m_{\tau}m_{p}}}{4(2\pi)^{2}s_{\tau\pk{K}}}
    \frac{|{\bf p}_{\pk{K}}|}{|{\bf p}_{\pk{d}}|} \pk{N_{I}}\,
    \frac{1}{6}\! \sum_{\rm spins} |\mathcal{M}|^{2},
\label{boundkin}
\end{equation}
where $\pk{m}_{i}$ and $p_{j}$ are the masses and momenta of the indicated 
particles
and $s_{\tau\pk{K}}$ is the c.m.\ energy squared. An isospin 
correction factor
$\pk{N_{I}}=9/4$ is included because of the contribution from
 $\pi^{0}$-exchange.

The matrix element is the integral over the Fermi momenta
\begin{equation}
    \mathcal{M} = \int\! \frac{\dkub{k}}{(2\pi)^{3}} 
    \frac{\dkub{q}}{(2\pi)^{3}} M,
\end{equation}
after the $q_{0}$ and $k_{0}$ integrations have been performed (giving a factor 
$\pk{m}/E\approx 1$ for each of the on-shell propagators when the Lorentz
boost is neglected). 

The expression for $M$ in the low energy limit and S-wave approximation becomes
\begin{equation}
    M = M_{\rm D}M_{\rm C}M_{\rm B}M_{\rm A}
    \frac{i}{p_{\pi}^{2}-\pk{m}_{\pi}^{2}+i\epsilon},
\end{equation}
where the respective vertex matrices are given by
\begin{eqnarray}
    M_{\rm D} & = & \eta_{\tau}^{\dagger} \left[ 
    (2\pi)^{\frac{3}{2}}\sqrt{2\pk{m_{d}}}
    \left( \frac{1}{\sqrt{3}} \ip{\bmath{\sigma}}{\bmath{\epsilon}_{\pk{d'}}}
    \right) \Psi_{\tau}^{\dagger}({\bf k})\right] \xi_{\Lambda} 
\label{Mdefd} \\
    M_{\rm C} & = & \xi_{\Lambda}^{\dagger} [\mathcal{G}] \xi_{\pk{n}} 
\label{Mdefc} \\
    M_{\rm B} & = & \xi_{\pk{n}}^{\dagger} 
    \left[ (2\pi)^{\frac{3}{2}}\sqrt{2\pk{m_{d}}}
    \left( \frac{-1}{\sqrt{2}}\ip{\bmath{\sigma}}{\bmath{\epsilon}_{\pk{d'}}}
    \right) \Phi_{\pk{d}}({\bf q}) \right] \xi_{\pk{p}^{c}} 
\label{Mdefb} \\
    M_{\rm A} & = & \xi_{\pk{p}^{c}}^{\dagger}
    [\mathcal{A}\ip{\hat{\bf p}}{\bmath{\epsilon}_{\pk{d'}}^{\dagger}}+
    i\mathcal{B}\ip{\bmath{\sigma}}
    {\cross{\bmath{\epsilon}_{\pk{d'}}^{\dagger}}{\hat{\bf p}}}]\eta_{\pk{p}}.
\label{Mdefa}
\end{eqnarray}
The parametrizations for the B and D vertices stem from vertex functions in the
pole
approximation, while for A, $\mathcal{A}$ and $\mathcal{B}$ are
invariant functions of the total energy only and for C, $\mathcal{G}$ is an
invariant function of energy and angle. Only S-wave contributions are 
considered
for the deuteron and the hypertriton.

Summation over internal spins and polarizations gives
\begin{eqnarray}
    \mathcal{M} & = &  \frac{2\pk{m_{d}}}{\sqrt{6}} \mathcal{G}
    \eta_{\tau}^{\dagger} [(\mathcal{A}-2\mathcal{B})
    \ip{\hat{\bf p}}{\bmath{\epsilon}_{\pk{d}}}+i\mathcal{A}\ip{\bmath{\sigma}}
    {\cross{\hat{\bf p}}{\bmath{\epsilon}_{\pk{d}}}}]\eta_{\pk{p}}
    G_{\pk{B}},
\end{eqnarray}
where the transition matrix $G_{\pk{B}}$ is defined as
\begin{equation}
    G_{\pk{B}} = i \int\! \frac{\dkub{q}\dkub{k}}{(2\pi)^{3}}
    \frac{\Psi_{\tau}^{\dagger}({\bf k}) \Phi_{\pk{d}}(\bf q)} 
    {({p^{0}_{\pi}})^{2}-{\bf p}_{\pi}^{2}-\pk{m}_{\pi}^{2}+i\epsilon}.
\end{equation}
The zeroth component of the virtual pion four-momentum is defined by the 
on-shell energies:
$p^{0}_{\pi}=E_{d}({\bf p}_{d})+E_{p}(-{\bf p}_{d})-
E_{d}(-(1\!-\!\gamma){\bf p}_{K}-{\bf k})
-E_{n}(\frac{1}{2}{\bf p}_{d}+{\bf k})$.
Since $\mathcal{B}$ is much smaller than $\mathcal{A}$~\cite{GW14}, it is
 possible to factorize, approximately, the total cross section in the forward
 direction as 
\begin{eqnarray}
    \lefteqn{ \left[ 
    \frac{d\sigma} {d\Omega} (\pk{pd}\rightarrow \nuclide{3}{\Lambda}{\! H}
    \pk{K}^{+}) \right]_{\pk{CM}} 
    = \frac{3\pk{m_{h}}s_{\!\Lambda \pk{K}}s_{\pi \pk{d}}}
    {\pk{m_{n}^{2}m_{\!\Lambda}}s_{\pk{hK}}} (2\pi)^{2} {|G_{\pk{B}}|}^{2}
    \frac{|{\bf p}_{\pk{K}}|} {|{\bf p}_{\pk{d}}|} } \nonumber \\ 
    & \times & \left[ \frac{|{\bf p}_{\pk{p}}|}{|{\bf p}_{\pi}|}
    \frac{d\sigma}{d\Omega}(\pk{pp'}\rightarrow \pi^{+}\pk{d}) 
    \right]_{\pk{cm}}^{\theta=0}
    \left[ \frac{|{\bf p}_{\pi}|}{|{\bf p}_{\pk{K}}|}
    \frac{d\sigma}{d\Omega}(\pi^{+}\pk{n}\rightarrow 
    \Lambda \pk{K}^{+}) \right]_{\pk{cm}}.
\label{xsboundtot}
\end{eqnarray}
The cross section for the vertex processes are given by
\begin{eqnarray}
    \left[ \frac{d\sigma}{d\Omega}(\pk{pp'} \rightarrow \pk{d}\pi^{+}) 
    \right]_{\pk{cm}} & = &
    \frac{\pk{m_{p}}^{2}} {8(2\pi)^{2}s_{\pi\pk{d}}} 
    \frac{|{\bf p}_{\pi}|}{|{\bf p}_{\pk{p}}|}
    |\mathcal{A}|^{2} \\
    \left[ \frac{d\sigma}{d\Omega}(\pi^{+}\pk{n} \rightarrow \Lambda\pk{K}^{+})
    \right]_{\pk{cm}} & = &
    \frac{\pk{m_{\Lambda}m_{n}}} {4(2\pi)^{2}s_{\!\Lambda\pk{K}}} 
    \frac{|{\bf p}_{\pk{K}}|}{|{\bf p}_{\pi}|} 
    |\mathcal{G}|^{2}.
\end{eqnarray}

\section{Break-up}
\label{sec:breakup}

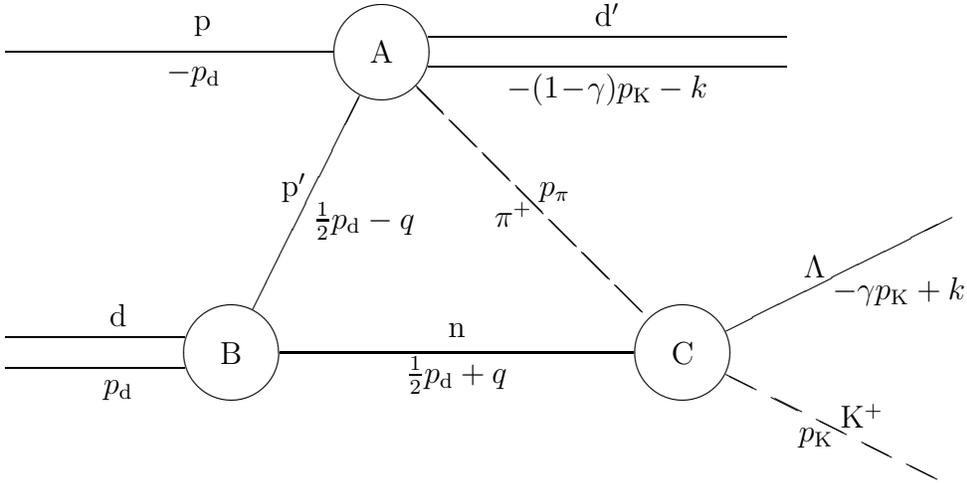
\begin{figure}[ht]

\begin{picture}(120,70)(-30,-20)


    \put(-30,-2){\line(1,0){23.81}}
        \put(-15,5){\makebox(0,0){$\pk{d}$}}
        \put(-15,-5){\makebox(0,0){$p_{\pk{d}}$}}
    \put(-30,2){\line(1,0){23.81}}
    \put(0,0){\circle{13}} \put(0,0){\makebox(0,0){B}}
    \put(6.5,0){\line(1,0){47}}
        \put(30,3){\makebox(0,0){$\pk{n}$}}
        \put(30,-3){\makebox(0,0){$\frac{1}{2}p_{\pk{d}}+q$}}
    \put(60,0){\circle{13}} \put(60,0){\makebox(0,0){C}}


    \put(-30,40){\line(1,0){43.5}}
        \put(-5,43){\makebox{$\pk{p}$}} 
        \put(-5,37){\makebox(0,0){$-p_{\pk{d}}$}}
    \put(20,40){\circle{13}} \put(20,40){\makebox(0,0){A}}
    \put(26.195,42){\line(1,0){47.61}}
        \put(50,45){\makebox(0,0){$\pk{d'}$}}
        \put(50,35){\makebox(0,0){$-(1\!-\!\gamma)p_{\pk{K}}-k$}}
    \put(26.195,38){\line(1,0){47.61}}


    \put(2.907,5.814){\line(1,2){14.186}}
        \put(10,20){\makebox(0,0)[br]{$\pk{p'}$}} 
        \put(11,20){\makebox(0,0)[tl]{$\frac{1}{2}p_{\pk{d}}-q$}}
    \multiput(24.596,35,404)(4.401,-4.401){7}{\line(1,-1){3.7}}
        \put(40,20){\makebox(0,0)[tr]{$\pi^{+}$}}
        \put(41,20){\makebox(0,0)[bl]{$p_{\pi}$}}
    \put(65.814,2.907){\line(2,1){30}}
        \put(79,10){\makebox(0,0)[br]{$\Lambda$}}
        \put(80,10){\makebox(0,0)[tl]{$-\gamma p_{\pk{K}}+k$}}
    \multiput(65.814,-2.907)(6,-3){5}{\line(2,-1){4}}
        \put(81,-10){\makebox(0,0)[bl]{$\pk{K}^{+}$}}
        \put(80,-10){\makebox(0,0)[tr]{$p_{\pk{K}}$}}

\end{picture}

    \caption{Feynman diagram of the break-up reaction,
    $\pk{pd} \rightarrow \pk{d \Lambda K}^{+}$,
    defining the various momenta. The partition of the $\pk{d}$ and $\Lambda$
    momenta
    is weighted by $\gamma = \pk{m_{\!\Lambda}/(m_{d}+m_{\!\Lambda})}=0.373$.
    The
    diagram for the $\pi^{0}$ case follows upon exchanging $\pk{p'}$ and
    $\pk{n}$.}
    \label{fig:feynbreakup}
\end{figure}

Because of the small binding energy of the hypertriton, there is a large
probability that the final deuteron and hyperon are scattered instead of forming
a bound state. This demands a calculation of the scattering cross section to be
compared with the bound state cross section.
The Feynman diagram for the break-up reaction is shown in 
Fig.~\ref{fig:feynbreakup}, where the kinematics is the same as before.

\subsection{Cross section}

The kinematic part of the cross section is calculated in the low energy
(non-relativistic) limit which gives
\begin{equation}
    \frac{d\sigma}{d\Omega} = \frac{\pk{m_{\tau}m_{p}}}
    {8(2\pi)^{5}\pk{m_{d}}s_{\tau\pk{K}}}
    \frac{1}{|{\bf p}_{\pk{d}}|} \pk{N_{I}} \int\! \dkub{P}\, |{\bf p}_{\pk{K}}|
    \, \frac{1}{6}\! \sum_{\rm spins} |\mathcal{M}|^{2},
\label{breakupkin}
\end{equation}
where $P$ is the relative momentum of the lambda-deuteron system and
\begin{equation}
    |{\bf p}_{\pk{K}}| = 
    \sqrt{2\mu_{\pk{K}\!-\!\Lambda \pk{d}}
    (Q-\frac{P^{2}}{2\mu_{\Lambda \pk{d}}})}
\end{equation}
is the momentum of the kaon in the final state.

The matrix element is now given, in the low energy limit, by
\begin{equation}
    \mathcal{M} = \int\! \frac{\dkub{q}}{(2\pi)^{3}} M,
\end{equation}
where the $q_{0}$-integration over the neutron propagator already is performed
and
\begin{equation}
    M = M_{\rm C}M_{\rm B}M_{\rm A}
    \frac{i}{p_{\pi}^{2}-\pk{m}_{\pi}^{2}+i\epsilon},
\label{Mbreakup}
\end{equation}
where the vertex matrices are given by Eqs~\refeq{Mdefc}--\refeq{Mdefa}.
Summing over internal spins and expanding the products of the Pauli matrices now
gives
\begin{equation}
    \mathcal{M} = -\mathcal{G} 
    \sqrt{\pk{m_{d}}} G_{\pk{S}} \eta_{\Lambda}^{\dagger}
    [-\mathcal{B} \ip{\hat{\bf p}}{\bmath{\epsilon}_{\pk{d}}}
    \ip{\bmath{\sigma}}{\bmath{\epsilon}_{\pk{d'}}^{\dagger}}+
    \ip{\bmath{\mathcal{Q}}}
    {\bmath{\epsilon}_{\pk{d'}}^{\dagger}}]\eta_{\pk{p}},
\label{Mextonly}
\end{equation}
where $G_{\pk{S}}$ is the transition matrix element
\begin{equation}
    G_{\pk{S}} = i\int\! \frac{\dkub{q}}{(2\pi)^{3/2}}  
    \frac{\Phi_{\pk{d}}({\bf q})}
    {({p^{0}_{\pi}})^{2}-{\bf p}_{\pi}^{2}-\pk{m}_{\pi}^{2}+i\epsilon}
\end{equation}
and the vector \bmath{\mathcal{Q}} is defined by
\begin{equation}
    \bmath{\mathcal{Q}} = \hat{\bf p} [\mathcal{A} 
    (\ip{\bmath{\sigma}}{\bmath{\epsilon}_{\pk{d}}})]+
    (\cross{\hat{\bf p}}{\bmath{\epsilon}_{\pk{d}}})[i\mathcal{B}]+
    \bmath{\epsilon}_{\pk{d}} [\mathcal{B}(\ip{\bmath{\sigma}}{\hat{\bf p}})].
\end{equation}
The spinor part of Eq.~\refeq{Mextonly} can be separated into two orthogonal
parts with the total spin of the deuteron-lambda system equal to 1/2 and 3/2
respectively. This is accomplished by using the projection operators
\begin{eqnarray}
    P'_{\frac{1}{2}} & = & \frac{1}{\sqrt{3}}
    \ip{\bmath{\sigma}}{\bmath{\epsilon}_{\pk{d'}}^{\dagger}}  \\
    P_{\frac{1}{2}} & = & \frac{1}{3}
    [\ip{\widehat{\bmath{\mathcal{Q}}}}{\bmath{\epsilon}_{\pk{d'}}^{\dagger}}-
    i\ip{\bmath{\sigma}}{\cross{\widehat{\bmath{\mathcal{Q}}}}
    {\bmath{\epsilon}_{\pk{d'}}^{\dagger}}}] \\
    P_{\frac{3}{2}} & = & \frac{1}{3}
    [2\ip{\widehat{\bmath{\mathcal{Q}}}}{\bmath{\epsilon}_{\pk{d'}}^{\dagger}}+
    i\ip{\bmath{\sigma}}{\cross{\widehat{\bmath{\mathcal{Q}}}}
    {\bmath{\epsilon}_{\pk{d'}}^{\dagger}}}],
\label{projdef}
\end{eqnarray} 
where the two spin-1/2 projections are orthogonal (after squaring and summing
 over external spins) to the spin-3/2 projection, but not to each other. 
If the $\mathcal{B}$ terms are neglected,  
the final form for the matrix element squared is
\begin{equation}
    \sum_{\rm spins} |\mathcal{M}|^{2} = 6 \pk{m_{d}}|\mathcal{AG}|^{2}
    (\frac{1}{3}|G_{\pk{S},\frac{1}{2}}|^{2}+
    \frac{2}{3}|G_{\pk{S},\frac{3}{2}}|^{2}).
\end{equation}

The total cross section in the forward direction becomes
\begin{eqnarray}
    \lefteqn{\hspace{-0.5cm} \left[ 
    \frac{d\sigma} {d\Omega}(\pk{pd}\rightarrow \pk{d}\Lambda \pk{K}^{+})
    \right]_{\pk{CM},s} = 
    \frac{3\pk{m_{\tau}}s_{\pi\pk{d}}} 
    {2\pi\pk{m_{n}^{2}m_{\!\Lambda}}s_{\tau\pk{K}}}
    \frac{3a_{s}} {|{\bf p}_{\pk{d}}|} 
    \left[ \frac{|{\bf p}_{\pk{p}}|} {|{\bf p}_{\pi}|}
    \frac{d\sigma}{d\Omega}(\pk{pp'}\rightarrow \pi^{+}\pk{d}) 
    \right]_{\pk{cm}}^{\theta=0} } \nonumber \\
    & \times & \int_{0}^{\sqrt{2\mu_{\!\Lambda \pk{d}}Q}} 
    \dkub{P} |{\bf p}_{\pk{K}}||G_{\pk{S},s}|^{2} s_{\!\Lambda \pk{K}}
    \left[ \frac{|{\bf p}_{\pi}|} {|{\bf p}_{\pk{K}}|}
    \frac{d\sigma}{d\Omega}(\pi^{+}\pk{n}\rightarrow \Lambda \pk{K}^{+})
    \right]_{\pk{cm}}.
\label{xsbreakuptot}
\end{eqnarray}
where the index $s$ denotes the different spin states $1/2$ and $3/2$ and
 $a_{s}$ the corresponding fractions $1/3$ and $2/3$.

\section{Approximation of transition matrices}
\label{sec:transmatr}

Introducing a new parameter $\kappa^{2}=(\kappa_{0})^{2}-\pk{m}_{\pi}^{2}$,
where $\kappa_{0}=p^{0}_{\pi}({\bf k}={\bf q}=0)$,
the pion propagator can be rewritten with the aid of the integral
\begin{equation}
\int\! \dkub{x}\, \expup{i\ip{{\bf p}_{\pi}\,}{\,\bf x}}
    \frac{\expup{i(\kappa +i\epsilon)r}}{r}
    = \frac{-4\pi}{\kappa^{2}-{\bf p}_{\pi}^{2}+i\epsilon}.
\end{equation}
Substituting this for the propagator in the transition matrices $G_{\pk{B}}$ and
$G_{\pk{S},s}$ they become (in configuration space)
\begin{eqnarray}
    G_{\pk{B}} & = & \frac{-i}{4\pi} \int\! \dkub{x}\,
    \psi_{\tau}^{\dagger}({\bf x})
    \frac{\expup{i\kappa r}}{r} \expup{i\ip{\bf p\,}{\,\bf x}} 
    \varphi_{\pk{d}}({\bf x}) \label{Gdefb} \\
    G_{\pk{S},s} & = & \frac{-i}{4\pi} \int\! \dkub{x}\, 1 
    \frac{\expup{i\kappa r}}{r} \expup{i\ip{\bf p\,}{\,\bf x}} 
    \varphi_{\pk{d}}({\bf x}),
\label{Gdefs}
\end{eqnarray}
where ${\bf p}$ is the pion momentum when Fermi momenta are neglected. 
The similarity of the expressions \refeq{Gdefb} and \refeq{Gdefs} suggests 
the way to include a scattering wave
function (not equal to unity) for the unbound lambda-deuteron system.
In order to do this properly, a bound state wave function related to the 
scattering state wave function is needed. A good and fairly simple
 candidate for the bound state is the wave function of Congleton~\cite{congl},
which, unfortunately, is not easily transformed to a scattering wave function.
A simpler alternative would be a two-term Hulth\'{e}n wave function, but its 
behaviour close to the origin is like $a-br$, while the Congleton function
goes like $a-cr^{2}$. This discrepancy could be remedied by using a
modified Hulth\'{e}n wave function (the result of using a
squared Yamaguchi form factor in a separable potential) of the type
\begin{equation}
    \psi_{\tau}(r) = N \frac{1}{r}\left[ \expup{-\alpha_{\tau}r}-
    \expup{-\beta_{\tau}r}-
    \frac{\beta_{\tau}^{2}+k^{2}}{2\beta_{\tau}} r\expup{-\beta_{\tau}r}
    \right]
\label{hulth}
\end{equation}
and then adjust the parameter $\beta_{\tau}$ to give the same rms radius as
Congleton.
 This leads to $\beta_{\tau} = 23.77 \alpha_{\tau}$ 
($\alpha_{\tau}=\sqrt{2\mu_{\pk{d\Lambda}}\pk{E_{B}}}=0.068\ \rm fm^{-1}$).
For the deuteron the S-wave part of the 
parametrized Paris wave function~\cite{paris} is employed.

\subsection{Scattering wave functions}

In scattering theory it is possible to relate the
scattering wave function to the bound state wave function. The expression for
this is deduced by F\"{a}ldt and Wilkin~\cite{FW1} and is given by
\begin{equation}
    \psi(r) = - \left[
    \sqrt{\frac{\alpha(\alpha^{2}+k^{2})}{2\pi}}\bar{\psi}^{(-)}(k,r)
    \right]_{k=i\alpha}.
\label{wfrel}
\end{equation}
 The scattering state wave function for the spin-1/2 $\Lambda$d-state
 has a form reminding of Eq.~\refeq{hulth}:
\begin{equation}
    \bar{\psi}_{\tau\frac{1}{2}}^{(-)}(k,r) =
    \frac{1}{kr} \left[ \cos{\delta}\sin{kr}+
    \sin{\delta}(\cos{kr}-\expup{-\beta_{\tau}r}-
    \frac{\beta_{\tau}^{2}+k^{2}}
    {2\beta_{\tau}}r\expup{-\beta_{\tau}r}) \right],
\end{equation}
where
\begin{equation}
    \tan{\delta}=
    \frac{-16\beta^{5}k(\alpha+\beta)^{4}}
    {(5\beta^{6}-15\beta^{4}k^{2}-5\beta^{2}k^{4}-k^{6})
    (\alpha+\beta)^{4}-
    (\beta^{2}+k^{2})^{4}
    ((2\beta+\alpha)^{2}+\beta^{2})}.
\end{equation}

\begin{figure}[h]
    \includegraphics{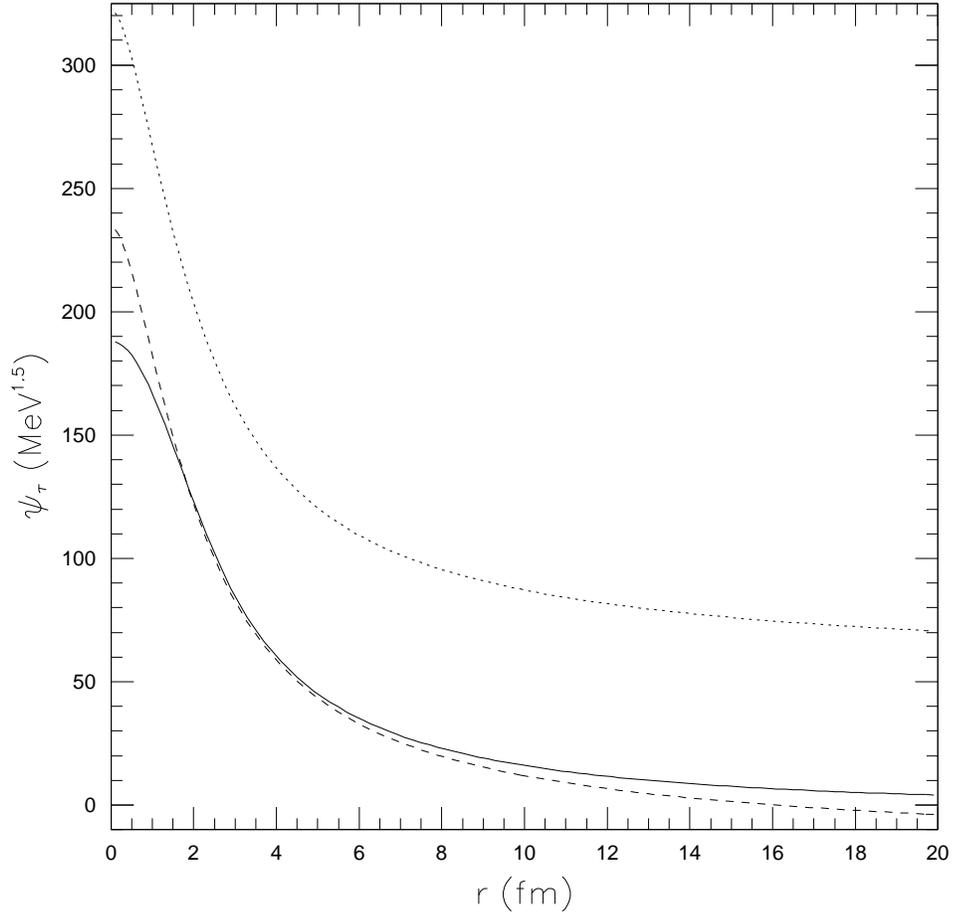}
    \caption{Comparison of the Congleton hypertriton wave function (solid line)
    with the
    spin-1/2 (dashed line) and spin-3/2 (dotted line) $\Lambda$d scattering
    wave functions. The scattering
    wave functions are normalized according to Eq.~\refeq{wfrel}, 
    which is approximatively valid even for $k\neq i\alpha$ (actually $k=0$ in
    this plot).}
\label{fig:wfplot}
\end{figure}

For the spin-3/2 $\Lambda$d-state, the same formul\ae\ are used but with 
another value for
$\alpha_{\tau}$. The new value is estimated from the expression for the
 potential strength 
\begin{equation}
    \lambda=-\frac{1}{4\pi N^{2}} \frac{\beta^{5}(\alpha+\beta)^{4}}
    {(2\beta+\alpha)^{2}+\beta^{2}},
\end{equation}
where $N$ is a suitable normalization constant (independent of $\alpha$).
It is stated in~\cite{mg95}
that the spin-3/2 case is unbound up to 1.32 times the strength parameter.
Keeping $\beta$ fixed, this gives an estimate for the strength parameter for a
 bound state
($\alpha=0$) which can be divided by 1.32 and used to calculate the new 
$\alpha'_{\tau}$ for the unbound state.
The calculations indicated above result in the values:
\[ \begin{array}{rcrl}
    \alpha_{\tau} & = & 0.068 & {\rm fm}^{-1} \\
    \beta_{\tau}  & = & 1.616 & {\rm fm}^{-1} \\
    \alpha'_{\tau} & = & -0.1334 & {\rm fm}^{-1}.
\end{array} \]
The wave functions for the bound state and the scattering states are plotted in 
Fig.~\ref{fig:wfplot}.

\subsection{Final expression for the transition matrices}

The angular integrations in Eqs~\refeq{Gdefb} and \refeq{Gdefs} can be
 performed in
 the S-wave approximation, giving (to first order),
\begin{eqnarray}
    G_{\pk{B}} & = & - \int\! rdr\, \psi_{\tau}^{\dagger}(r)
    \frac{\expup{i\kappa r}}{r} j_{0}((1-\gamma)p_{\pk{K}}r)
    j_{0}(\frac{1}{2}p_{\pk{d}}r) \varphi_{\pk{d}}(r) \\
    G_{\pk{S},s} & = & - \int\! rdr\, {\psi_{\tau,s}^{(-)}}^{\dagger}(k,r)
    \frac{\expup{i\kappa r}}{r} j_{0}((1-\gamma)p_{\pk{K}}r)
    j_{0}(\frac{1}{2}p_{\pk{d}}r) \varphi_{\pk{d}}(r).
\end{eqnarray}

\section{Parametrization of vertices}
\label{sec:vertpara}

The cross sections for the vertices A and C, which turn up on the right hand
sides of 
Eqs~\refeq{xsboundtot} and \refeq{xsbreakuptot}, are parametrized empirically.

\subsection{Empirical fit of $\pk{pp}\rightarrow \pi^{+}\pk{d}$ data}
\label{sec:cwfit}

F\"{a}ldt and Wilkin~\cite{FW2} use cross section data for the reaction 
$\pi^{+}\pk{d}\rightarrow \pk{pp}$ and fit them to an exponential form. 
By the principle of detailed balance the $\pk{pp}\rightarrow \pi^{+}\pk{d}$ and 
$\pi^{+}\pk{d}\rightarrow \pk{pp}$ reactions have a simple relationship:
\begin{eqnarray}
    \lefteqn{ \left[ \frac{|{\bf p}_{\pk{p}}|}{|{\bf p}_{\pi}|}
    \frac{d\sigma}{d\Omega}(\pk{pp}\rightarrow \pi \pk{d})
    \right]_{\pk{cm}}^{\theta=0}}    \nonumber \\
    & = & \frac{3}{4} \left[ \frac{|{\bf p}_{\pi}|}{|{\bf p}_{\pk{p}}|}
    \frac{d\sigma}{d\Omega}(\pi \pk{d}\rightarrow \pk{pp}) 
    \right]_{\pk{cm}}^{\theta=0}    \nonumber \\
    & = & \frac{3}{4} \left[ \frac{|{\bf p}_{\pi}|}{|{\bf p}_{\pk{p}}|}
    \right]_{\pk{cm}}
    \left\{ \expup{2.5914-0.011115T_{\pi}}+0.000065(T_{\pi}-500) \right\},
\label{xsAfit}
\end{eqnarray}
where the exponential fit is inserted in the last step. In this expression
$T_{\pi}$ is the laboratory kinetic energy of the pion (in MeV) in the reaction
$\pi^{+}\pk{d}\rightarrow \pk{pp}$, giving the cross section in mb/sr.
The threshold for the $\pk{pd}\rightarrow\nuclide{3}{\Lambda}{\! H}\pk{K}^{+}$
reaction corresponds to $T_{\pi}=419.5$ MeV, 
$|{\bf p}_{\pi}|=428$ MeV/c and $|{\bf p}_{\pk{p}}|=727$ MeV/c, yielding a 
cross section of 53.3 $\mu$b/sr according to Eq.~\refeq{xsAfit}. The Fermi
 momenta are neglected in this calculation.

\subsection{Empirical fit of $\pi^{+}\pk{n}\rightarrow \Lambda \pk{K}^{+}$ data}

There is a recent fit of the cross section for the reaction
$\pi^{+}\pk{n}\rightarrow \Lambda \pk{K}^{+}$ to a simple
formula \cite{cugnon}, but this does not have the correct energy dependence 
near threshold, which is crucial for the present application. Instead the
 threshold fit of Jones 
\emph{et al.}~\cite{jones} is used. 
This gives
\begin{eqnarray} 
    \frac{d\sigma}{d\Omega}(\pi^{+}\pk{n}\rightarrow\Lambda\pk{K}^{+}) =
    \frac{1} {4\pi} \left[ A {p}_{\pk{K}}+ 
    \frac{B{p}_{\pk{K}}^{3}} {1+(R{p}_{\pk{K}})^{2}} \right] & &
    p_{\pi}^{\rm lab}\! < 970 \label{xsCfita} 
\end{eqnarray}
where $A = 122\cdot 10^{-6}\ {\rm fm^{2}/MeV}$,
$B = 12\cdot 10^{-9}\ {\rm fm^{2}/MeV}$ and
$R = 7.165\cdot 10^{-3}\ {\rm (MeV)^{-1}}$ (the Compton wavelength of the pion).
At threshold the amplitude squared to use in Eqs~\refeq{xsboundtot} and
\refeq{xsbreakuptot} is 50.3 $\mu$b/sr.

\section{Results and conclusions}
\label{sec:numres}

\begin{figure}[h]
    \includegraphics{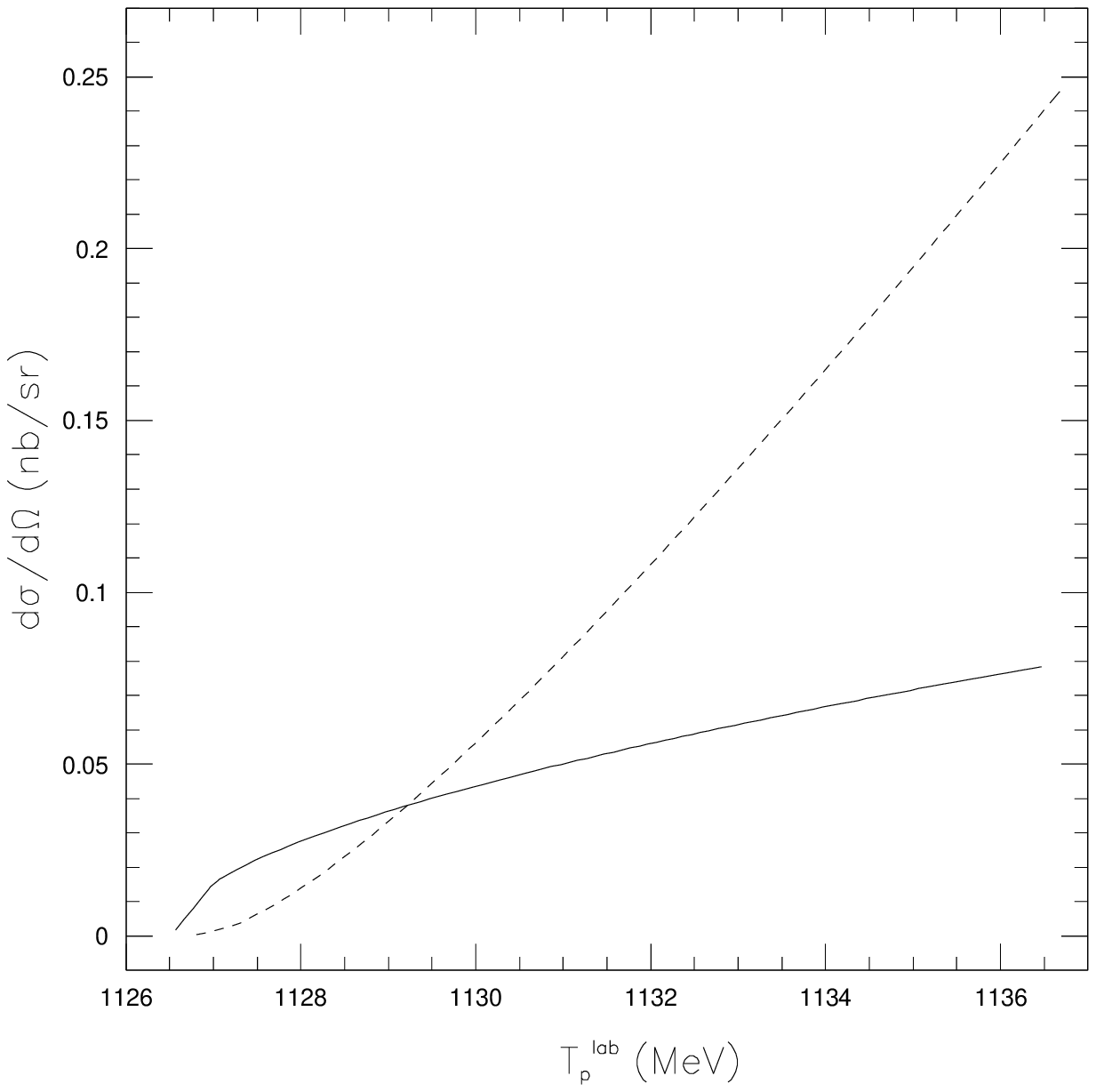}
    \caption{The differential cross sections for the reactions 
    $\pk{pd} \rightarrow \protect\nuclide{3}{\Lambda}{\! H}\pk{K}^{+}$ 
    (solid line) and 
    $\pk{pd} \rightarrow \pk{d \Lambda K}^{+}$
    (dashed line) in the forward direction.}
\label{fig:xsplot}
\end{figure}

The differential cross sections for the two processes 
$\pk{pd}\rightarrow\nuclide{3}{\Lambda}{\! H}\pk{K}^{+}$ and 
$\pk{pd}\rightarrow \pk{d}\Lambda\pk{K}^{+}$ have been calculated in the
 forward direction for energies up to $T_{\pk{p}}^{\rm lab}=1137\ {\rm MeV}$,
\emph{i.e.}, 10 MeV
above threshold. The wave functions employed are discussed in 
Sec.~\ref{sec:transmatr}. The result of the calculation is plotted in
 Fig.~\ref{fig:xsplot}. 
In addition the threshold amplitude squared 
($|f|^{2} =p_{\pk{in}}/p_{\pk{fin}} d\sigma /d\Omega $)
 for the reaction leading to a bound final state has been computed and found
 to be
\[
    |f(\pk{pd}\rightarrow\nuclide{3}{\Lambda}{\! H}\pk{K}^{+})|^{2} =
    1.03\ {\rm nb/sr}.
\]

The cross section for the 
$\pk{pd}\rightarrow\nuclide{3}{\Lambda}{\! H}\pk{K}^{+}$ reaction
is small and in addition almost completely drowned by the 
break-up reaction $\pk{pd}\rightarrow \pk{d}\Lambda\pk{K}^{+}$. Thus, if one has
the ambition to study a bound final state, it is of
utmost importance to identify the hypertriton.
The ratio of the cross sections for the bound and unbound final states in
the $\Lambda$d spin-1/2 channel is given by~\cite{FW1}
\begin{equation}
    \frac{\sigma_{\frac{1}{2}}(\pk{d\Lambda K})}
    {\sigma (\nuclide{3}{\Lambda}{\! H}\pk{K})} \approx
    \frac{1}{4} \left( \frac{Q}{\pk{E_{B}}} \right)^{\frac{3}{2}} 
    \left( 1+\sqrt{1+Q/\pk{E_{B}}} \right)^{-2},
\label{qratio}
\end{equation}
where the $Q$-value refers to the reaction 
$\pk{pd}\rightarrow\pk{d\Lambda K^{+}}$.
It is evident from this formula that the break-up reaction will dominate when 
$Q$ is greater than $\pk{E_{B}}$ and that this will happen at small $Q$
 because of the small binding energy ($\pk{E_{B}}= 0.13$ MeV). 
The ratio of the cross sections calculated from Eqs~\refeq{xsbreakuptot} and 
\refeq{xsboundtot} for the $\Lambda$d spin-1/2 case agrees 
 with Eq.~\refeq{qratio}. 
The cross section for the unbound $\Lambda$d spin-3/2 channel is
 larger than the cross section for the unbound $\Lambda$d spin-1/2 channel, by 
a factor
0.7 near threshold and increasing to 1.9 at $T_{\pk{p}}^{\rm lab} = 1137$ MeV.

Some future improvements should include more accurate wave functions (like
the one of Miyagawa \emph{et al.}~\cite{mg95}), higher energies and angular
dependence. The deuteron wave function should be corrected for the Lorentz
 boost.
 It would also be valuable to have a better experimental determination of the 
$\pi^{+} \pk{n}\rightarrow \Lambda \pk{K}^{+}$ vertex.

\section*{Acknowledgement}

I would like to thank my supervisor G\"{o}ran F\"{a}ldt and his colleague 
Colin Wilkin for valuable advice and patient endurance.

\newpage


\end{document}